\documentclass[12pt]{article}
\usepackage{epsfig}

\newcommand{\eq}[1]{(\ref{#1})}
\newcommand{\diff}{\partial}
\newcommand{\beq}{\begin{equation}}
\newcommand{\eeq}{\end{equation}}
\newcommand{\beqn}{\begin{eqnarray}}
\newcommand{\eeqn}{\end{eqnarray}}
\newcommand{\C}{{\cal C}}
\newcommand{\arctg}{\mathop{\rm arctg}}
\newcommand{\dd}{{\mathrm d}}

\begin{document}

~\vspace{-5mm}
\begin{flushright}   {\large ITEP-TH-75/99}  \end{flushright}
\vspace{15mm}

\centerline{\bf{SHORT STRINGS AND NEW PHYSICS PERSPECTIVES IN QCD
\footnote{Talk given by M.I.~Polikarpov at the workshop
"Lattice Fermions and Structure of the Vacuum",
5-9 October, Dubna, Russia.}}}

\vspace{10mm}
\centerline{{\bf M.N. Chernodub, M.I. Polikarpov}}
\vspace{3mm}
\centerline{Institute of Theoretical and Experimental Physics,}
\centerline{
B. Cheremushkinskaya, 25, 
117259 Moscow}
\vspace{5mm} 
\centerline{{\bf F.V. Gubarev, V.I. Zakharov}}
\vspace{3mm}
\centerline{Max-Planck Institut f\"ur Physik,}
\centerline{F\"ohringer Ring 6, 80805 M\"unchen, Germany}
\vspace{10mm}

\begin{abstract}
Within the (dual) Abelian Higgs model,
confining strings do not disappear at small distances but rather become
short strings. In compact 3D electrodynamics, as we argue here,
the confining strings are also manifested at small distances
in unusual power corrections, disobeying the standard rules of
the Operator Product expansion. 
In the most interesting case of QCD, there is yet no
derivation of short strings and we turn to phenomenology to
find evidence pro or contra their existence. The short strings in
QCD lead to non-standard power corrections. A
tentative conclusion 
of the analysis of existing data,
both in numerical simulations and in real experiments,
is that the novel power corrections are present,
at least at the mass scale of $(1\div 2)$ GeV.
\end{abstract}

\vspace{-2mm}

\section{Introduction.}
The strings which are responsible for confinement in the infrared
region can survive \cite{ShoStr2} in the ultraviolet region as well and
be responsible for non-perturbative effects at small
distances. The simplest manifestation of the short strings
in QCD would be a stringy piece in the heavy quark--antiquark
potential at {\it small} distances:
\beq\label{SS}
V(r) = -\alpha/r + \sigma_0 r, \,\, r \to 0
\eeq
This linear piece could be related
to divergences of the perturbative series in large orders revealed by
the so called ultraviolet renormalon 
(see \cite{ShoStr1} and references therein).
In this sense, one can speculate that 
the short strings is  
a non-perturbative counterpart of the ultraviolet renormalon. 

It is of course far from being trivial to find the non-perturbative
potential at short distances in QCD. Thus, we are invited to consider
simpler model with confinement. So far, only the Abelian Higgs model
has been analyzed and the stringy potential at short distances was
indeed found \cite{ShoStr2}.  The physics behind the stringy potential
is highly non-trivial and can be viewed as a manifestation of the
Dirac strings. It is worth mentioning that the physical
manifestations of the Dirac strings were found first in the example of
the compact photodynamics \cite{polyakov}. In Sect. 2 we review the
results obtained in the case of the Abelian Higgs model
and comment on the connections with the compact $U(1)$. 
In Sect. 3 we
consider the potential at short distances within another $U(1)$ model,
namely the compact 3D electrodynamics. As is well known, it exhibits
confinement of the electric charges \cite{U1}, i.e. the linear
potential at large distances.
We do find a non-analytic behavior of the potential
at short distances. However, as is argued in Sect. 4, 
the new non-analytical terms may disappear
once the distance $r$ is much smaller than the size of monopoles
present in the model.
All the consideration here is on the classical level.

In case of QCD the use of the lattice regularization assumes that the
Dirac strings are allowed and carry no action.  From this point of
view the situation is a reminiscent of the Abelian models mentioned
above. However, unlike the abelian case
the monopoles associated with the end points of the
Dirac string may have zero action.
Thus,  
both classical field configurations and the quantum
running of the effective coupling seem to be equally important in the
QCD case.  As a result, there is no definite prediction for the short
distance behavior of the potential
at the moment. We concentrate therefore, on
phenomenological manifestations of the hypothetical short strings.  A
comparison with existing data indicates that the novel effects 
corresponding to
the short strings are indeed present. Naturally enough, the data
refer to a limited range of distances.  Thus, the statement above
refers to distances of order $(0.5~\div~1.0)\mbox{GeV}^{-1}$.  The QCD
phenomenology is reviewed in Sects. 5,6 while in Sect. 7 
conclusions are given.

\section{Short Strings in the Dual Abelian Higgs Model.}

The first example of drastic  
non-perturbative effects in ultraviolet
was in fact given in paper \cite{polyakov}.
The Lagrangian considered is that of free photons:
\beq
L~=~\frac{1}{4e^2} F^2_{\mu\nu}
\eeq
where $F_{\mu\nu}$ is the field strength tensor of the electromagnetic field.
Although the theory looks absolutely trivial, it is not the case
if one admits the Dirac strings.
Naively, the energy associated with the Dirac strings is infinite:
\beq
E_{\mbox{\small Dirac string}}={1\over 8\pi}\int d^3r\,{\bf H}^2\sim l\cdot A
\left({\mbox{magnetic~flux}\over A}\right)^2\to \infty
\eeq
where $l, A$ are the length and area of the string, respectively.
Since the magnetic flux carried by the string is quantized and finite
the energy diverges quadratically in the ultraviolet, i.e. in the
limit $A\to 0$.  
However within the lattice regularization the action 
of the string is in fact zero
because of the compactness of the $U(1)$.
The invisible Dirac strings may end up with monopoles
which have a non-zero action. Moreover, the monopole action is
linearly divergent in ultraviolet.
However the balance between the suppression due
to this action and enhancement due to the entropy
factor favors
a phase
transition to the monopole condensation at $e^2\sim 1$. 
As a result the test electric charges are subject
to linear potential at all the distances if $e^2$ is large enough.

Thus, in compact $U(1)$
 model the non-perturbative effects change the interaction
at all distances, for a range of the coupling values.
Next, one considers the Dual Abelian Higgs Model 
with the action
\beq
\label{AHM_action} S= \int d^4x \left\{ \frac{1}{4g^2}
F^2_{\mu\nu} + \frac{1}{2} |(\diff - i A)\Phi|^2 + \frac{1}{4} \lambda
(|\Phi|^2-\eta^2)^2 \right\}, 
\eeq
here $g$ is the magnetic charge,
$F_{\mu\nu}\equiv\diff_{\mu}A_{\nu}-\diff_{\nu}A_{\mu}$. The gauge boson and
the Higgs are massive, $m^2_V=g^2\eta^2, m_H^2= 2
\lambda \eta^2$.  
There is a well known Abrikosov-Nielsen-Olesen (ANO) solution to
the corresponding equations of motion. 
The dual ANO string may end up with electric charges.
As a result, the potential
for a test charge-anticharge pair grows linearly at large  distances:
\beq
V(r) = \sigma_\infty r, \,\, r \to \infty.
\eeq
Note that there is a Dirac string resting along the axis
of the ANO string connecting monopoles 
and its energy is still normalized to zero.

An amusing effect occurs if one goes to distances much smaller than
the characteristic mass scales $m_{V,H}^{-1}$. Then the ANO string
is peeled off and one deals with a naked (dual) Dirac string.
The manifestation of the string is that the Higgs field has to vanish
along a line connecting the external charges. Otherwise, 
the energy of the Dirac string would jump to infinity anew.  

As a result of the boundary condition that $\Phi$ vanishes on a line
connecting the charges, the potential contains a stringy piece \eq{SS}
at short distances \cite{ShoStr2}.
The string tension $\sigma_0$ smoothly depends on the ratio
$m_H/m_V$.
In particular, in the Bogomolny limit ($m_H = m_V$) the string tension
\beq
\sigma_0~\approx~\sigma_\infty,
\eeq
i.e. the effective string tension is the same at all distances.

\section{Short Strings in 3D Compact Electrodynamics.}

As it is well known \cite{U1} in 3D compact electrodynamics the
charge--anticharge potential is linear at large separations. Below we
consider the string tension $\sigma_0$ at small distances and show that
it has a non-analytical piece associated with small distances.

As usual, it is
convenient to perform the duality transformation, and work with the
corresponding
Sine-Gordon theory. The expectation value of the Wilson loop in dual
variables is:
\beq\label{Schi-1}
W = \frac{1}{\cal Z} \int {\cal D}\chi e^{- S(\chi,\eta_\C)},
\eeq
where 
\beq \label{Schi}
S(\chi,\eta_\C) = \left({e\over 2\pi}\right)^2
\int\, d^3 x
\left\{
{1\over 2}(\vec{\diff} \chi)^2 + m_D^2 (1-\cos[\chi-\eta_\C])
\right\},
\eeq
$m_D$ is the Debye mass and $S(\chi,0)$ is the action of the model. If
static charge and anticharge are placed at the points $(-R/2,0)$ and
$(R/2,0)$ in the $x_1, x_2$ plane ($x_3$ is the time axis), then
\beq \label{etaC}
\eta_\C = \arctg[{x_2 \over x_1-R/2}] - \arctg[{x_2 \over x_1+R/2}],
\,\, -\pi \leq \eta_\C \leq \pi.
\eeq

Below we present the results of the numerical calculations of the
string tension,
\beqn
\sigma = \partial E/ \partial (m_D R)\,\, , \label{sigma}\\
E = \int d^2 x
\left\{
{1\over 2}(\vec{\diff} \chi)^2 + m_D^2 (1-\cos[\chi-\eta_\C])
\right\}\,\, . \label{energy}
\eeqn
Note that the energy $E$ is measured now in units of the
dimensional factor $(\frac{e}{2\pi})^2$ ({\it cf.}
\eq{Schi}). Variation of functional \eq{energy} leads to the equation
of motion $\Delta \chi = m^2_D \sin[\chi - \eta_\C]$. 
For finite $R$ we can solve this equation numerically.
The energy $E$ versus $m_D R$ is shown on
Fig.1(a). At large separations between 
the charges ($m_D R \gg 1$) it tends to the asymptotic linear
behavior $E = 8 m_D R$ which can be obtained also analytically \cite{U1}). 

At small distances there is a contributions of Yukawa-type to the 
energy (\ref{energy}), which should be
extracted explicitly. Note that in course of rewriting original $3D$ compact
electrodynamics in the form (\ref{Schi-1}-\ref{Schi}) the Coulomb potential was
already subtracted, so that (\ref{energy}) contains Yukawa-like piece without singularity
at $R=0$. It is not difficult to find the corresponding coefficient:
\beq
\label{energy-1}
E = E^{string} - 2\pi ( K_0[m_DR]+\ln[m_D R] )
\eeq
where $K_0(x)$ is the modified Bessel function and $E^{string}$
is the energy of the charge--anticharge pair. 
The corresponding string tension
\beq\label{sigma-1}
\sigma^{string} = \sigma + 2\pi ( - K_1[m_DR]+ { 1\over m_D R} )
\eeq
is shown on Fig.1(b).  We found that the best fit of numerical data
for small values of $m_D R$ is by the function  
$\sigma^{string} = const \cdot (m_DR)^\nu$ which gives $\nu \approx 0.6$.

\begin{figure}
 \begin{minipage}{12.5cm} \begin{center}
 \epsfig{file=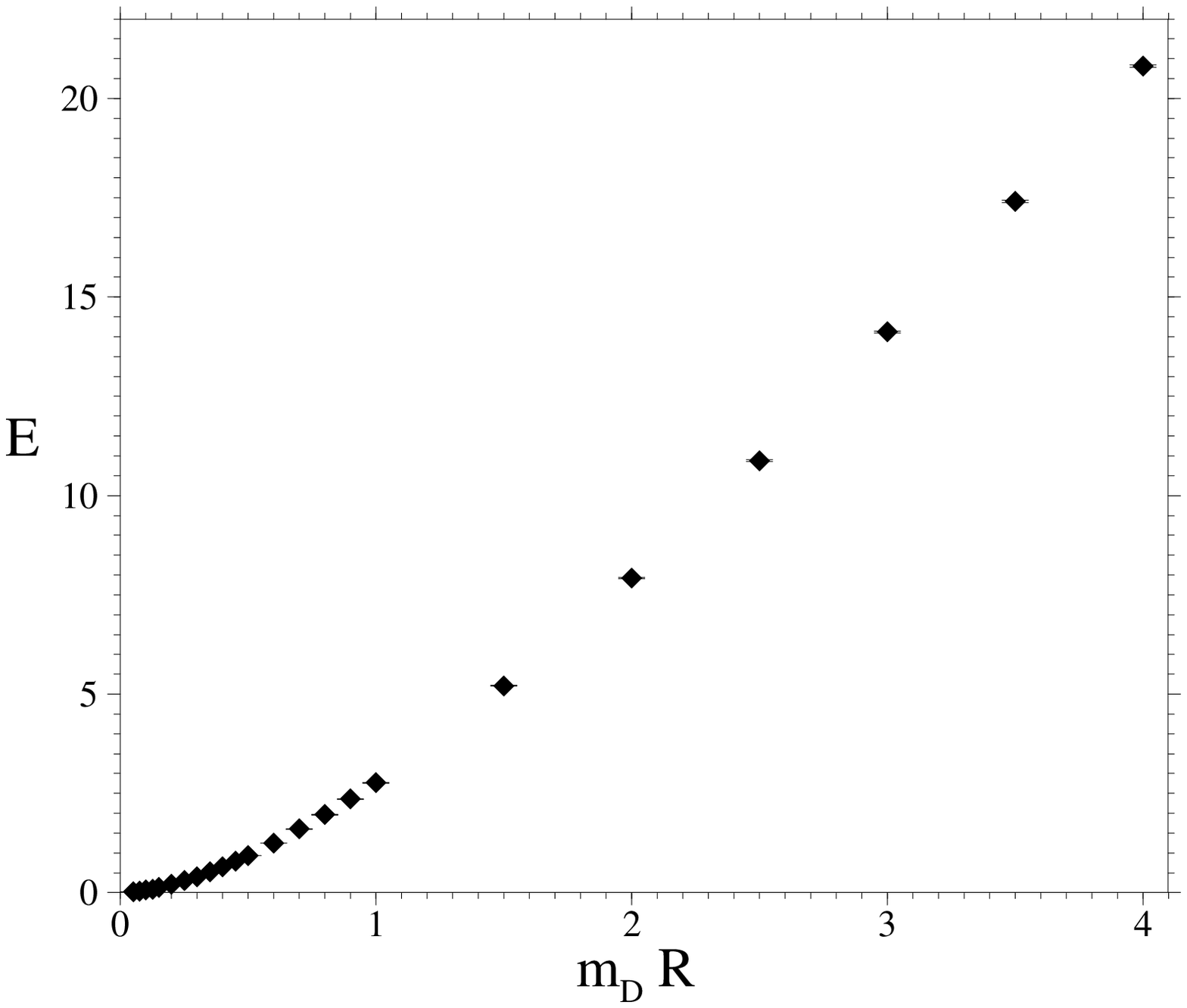,width=5.5cm,height=5.0cm} \hspace{6mm}
 \epsfig{file=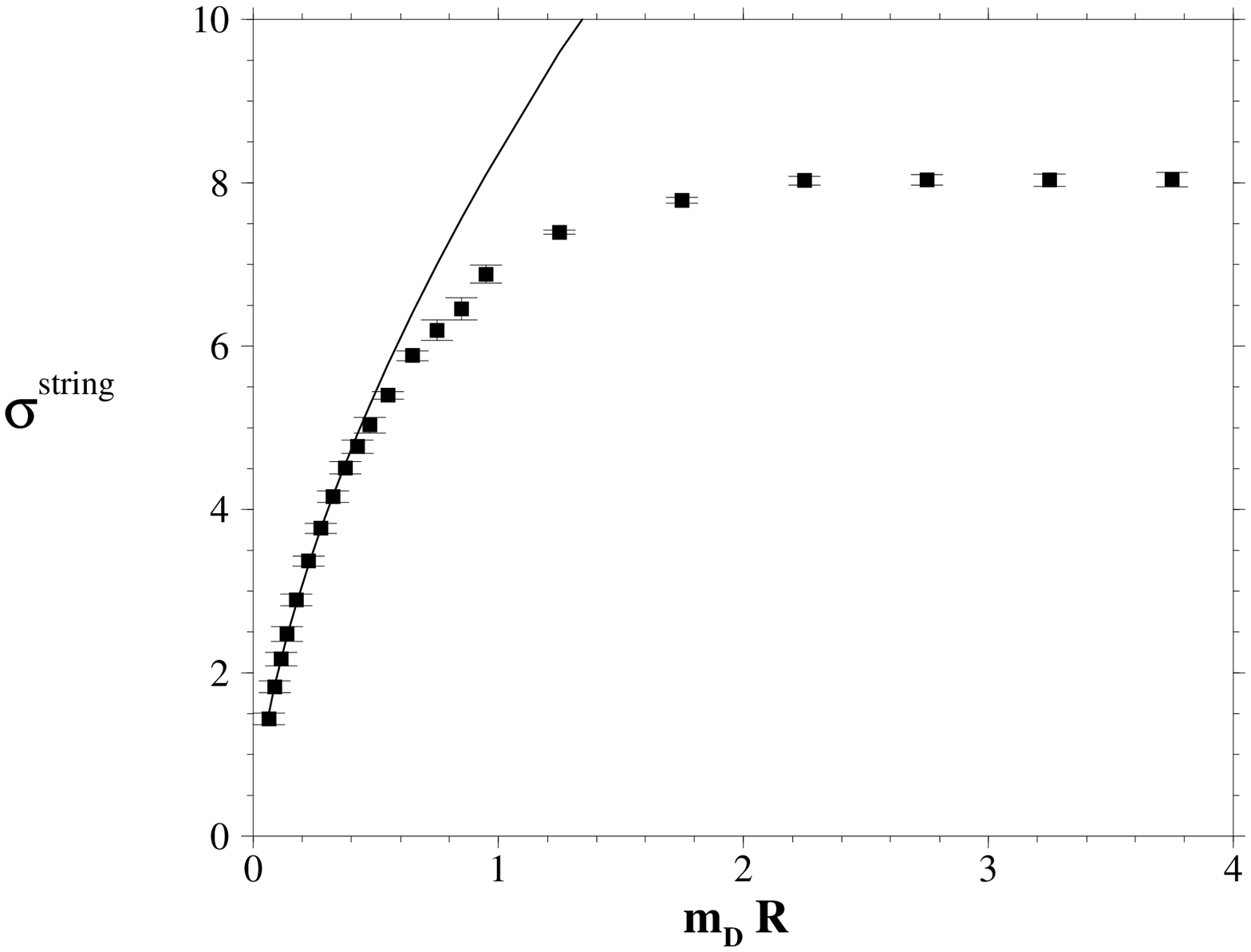,width=5.5cm,height=5.0cm}\\ (a) \hspace{5.5cm}
 (b)\\ \end{center} \end{minipage} \caption{
(a) The string tension (\ref{sigma}) of the charge -- anticharge 
separated by the distance $R$  in 3D compact U(1)theory;
(b) The string tension $\sigma^{string}$ (\ref{sigma-1}) as a function 
of $m_D R$ with corresponding best fitting function (see text).}
\label{1}
\end{figure} 

Thus the non-analytical potential associated with small distances
is softer than in the case of the Abelian Higgs model.
The source of the non-analyticity is the behavior of the function
$\eta_\C(x_1,x_2)$ \eq{etaC} 
which is singular
along the line connecting the charges, see Fig.2(a).

\section{Georgi-Glashow model}

The compact electrodynamics is usually considered as the limit of
Georgi--Glashow model, when the radius of the 't Hooft -- Polyakov
monopole tends to zero. For a non-vanishing monopole size the problem of
evaluating the potential at small distances becomes
rather complicated. To avoid unnecessary 
further complications we consider
the 3D Georgi--Glashow model in the BPS limit. The 't Hooft
Polyakov monopole corresponds then to the fields:
\beqn
\Phi^a & = & \frac{x^a}{r} \, \Biggl(\frac{1}{\tanh (\mu r)} - 
\frac{1}{\mu r}\Biggr)\,,\\
A^a_i & = & - \varepsilon^{aic} \frac{x^a}{r} \, \Biggl(\frac{1}{r} -
\frac{\mu}{\sinh(\mu r)} \Biggr)\, , \quad A^a_0 = 0\,.
\eeqn
The contribution of this monopole to the 
{\it full non-Abelian} Wilson loop $W$ can be calculated analytically.
If the
static charges are placed at points $\pm \vec{R}/2$ in the
$x_1,x_2$ plane the result is:
\beq \label{Wmon}
W({\vec b}_1,{\vec b}_2,\mu) = \cos h(\mu b_1) \, \cos h(\mu b_2) + 
\frac{({\vec b}_1 \cdot {\vec b}_2)}{b_1\,b_2} 
\sin h(\mu b_1) \, \sin h(\mu b_2)\,,
\label{WLeq}
\eeq
here ${\vec b}_{1,2} = \vec{x}_0 \pm {\vec R} \slash 2$, $b_k = |{\vec
b}_k|$, $\vec{x}_0$ is the center of the 't Hooft -- Polyakov monopole
and
\beq
h(x) = \frac{\pi}{2} - 
\frac{x}{2} \, \int\nolimits^{+\infty}_{- \infty} 
\frac{\dd \zeta}{\sqrt{x^2 + \zeta^2}
\sinh \sqrt{x^2 + \zeta^2}}\, .
\eeq
One way to represent \eq{Wmon} in terms of the function
$\eta_\C$ introduced earlier is:
\beqn \label{etaW}
\eta_\C(x_0,R,\mu) = 
{\mathrm {sign}}(y) \, \arccos W({\vec b}_1,{\vec b}_2,\mu)\, .
\eeqn
In the limit $R \mu \to \infty$ $W({\vec b}_1,{\vec b}_2,\mu) \to \cos
\eta_\C$ and $\eta_\C(x_0,R,\mu)$ coincides with the definition \eq{etaC}.
For small $R \mu$, $\eta_\C$ is singular not only between external
charges, but also outside this region (see Fig.2(b))
although the strength of singularity gets smaller.
In the limit of vanishing
$\eta_\C$ $\sigma_0$ should vanish.

\begin{figure}
 \begin{minipage}{12.5cm} \begin{center}
 \epsfig{file=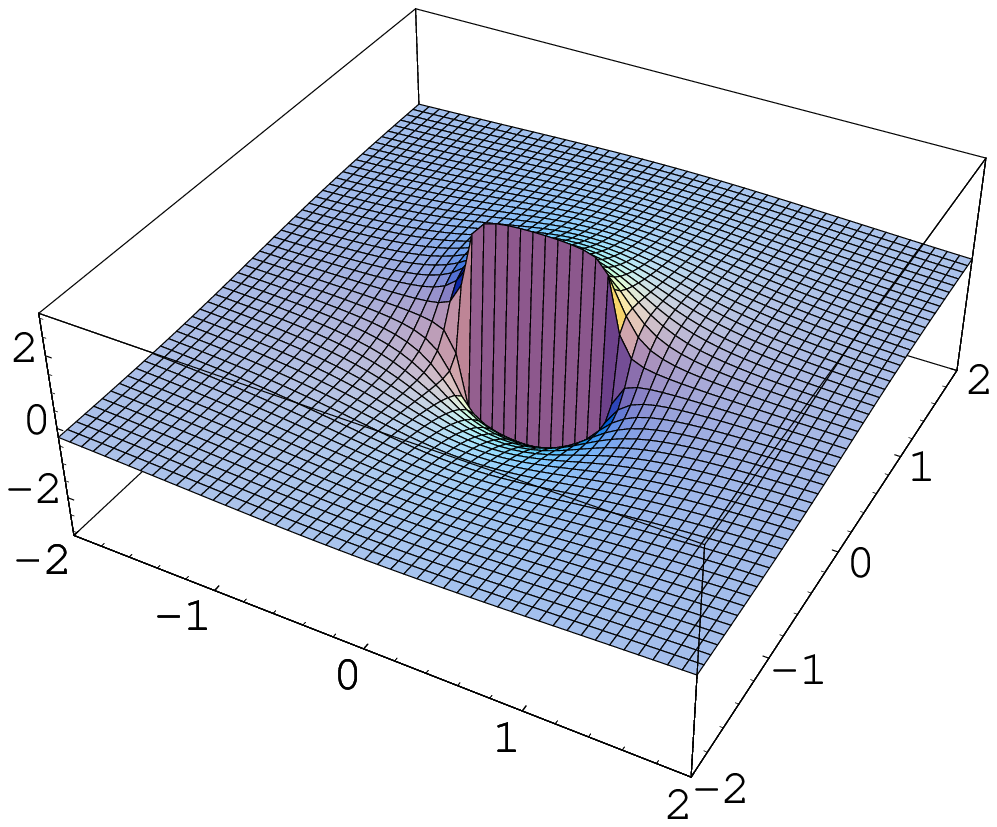,width=5.5cm,height=5.0cm} \hspace{6mm}
 \epsfig{file=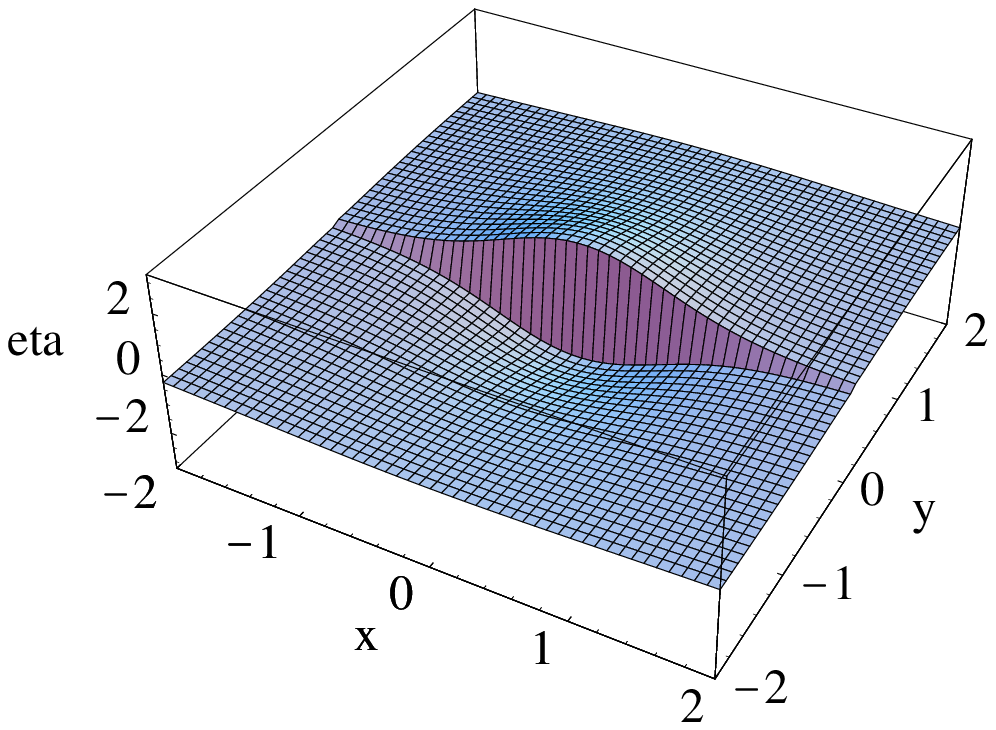,width=5.5cm,height=5.0cm}\\ (a) \hspace{5.5cm}
 (b)\\ \end{center} \end{minipage} \caption{Function $\eta_\C$ in
 compact $U(1)$ theory (a) and for extended monopoles (b),
 eq. \eq{etaW}}
\end{figure}

To summarize, it is natural to expect that at distances much smaller
than the monopole size, the non-analytical piece in the potential
associated with small distances disappears.  However, for a consistent
treatment of the problem one should take into account the
modification of the Coulomb-like monopole interaction due to the
finite size of the BPS monopoles.

\section{Topological defects and short-distance potential in QCD.}

Knowing the physics of the Abelian models above it is easy
to argue that the perturbative vacuum of QCD is not stable in fact. Indeed,
let us make the lattice more coarser
a la Wilson until the effective coupling of QCD would reach the
value where the phase transition in the compact $U(1)$ occurs. 
Then the QCD perturbative vacuum is unstable against the monopole formation.
The actual non-perturbative vacuum can of course be very different
but it cannot remain perturbative. Similar remark with respect to
formation of $Z_2$ vortices was in fact made long time ago \cite{mack}.

Thus, it is natural to expect that singular non-perturbative defects
play a role in QCD as well. In case of the abelian projection these
are Dirac strings with monopoles at the ends, while in case of the
$Z_2$ vortices the corresponding infinitely thin objects can be
identified with the so called P-vortices, see \cite{Deb97} and
references therein.

The existence of the infinitely thin topological defects 
in QCD makes it close 
akin of the Abelian models considered above. 
However, the non-Abelian nature of the interaction
brings in an important difference as well. 
Namely, the topological defects in 
QCD are marked rather by singular potentials
than by a large non-Abelian action. Consider first the Dirac string. 

Introduce to this end a potential which is a pure gauge:
\beq
A_{\mu}~=~\Omega^{-1}\partial_{\mu}\Omega\label{leer}
\eeq
and choose the matrix $\Omega$ in the form:
\beq
\Omega(x)~=~\left(\matrix{\cos{\gamma\over 2}& \sin{\gamma\over 2}e^{-i\alpha}\cr
- \sin{\gamma\over 2}e^{i\alpha}& \cos{\gamma\over 2}\cr}\right)
\eeq
where $\alpha$ and $\gamma$ are azimuthal and polar angles, respectively.
Then it is straightforward to check that
we generated a Dirac string directed along the $x_3$-axis
ending at $x_3=0$ and carrying the color index $a=3$.
It is quite obvious that such Abelian-like strings are allowed by
the lattice regularization of the theory. 

The crucial point, however, is that the non-Abelian action associated
with the potential \eq{leer} is identical zero. On the other hand, in
its Abelian components the potential looks as a Dirac monopole, which
are known to play important role in the Abelian projection of QCD (for
a review see, e.g., \cite{chernodub98}).  Thus, there is a kind of
mismatch between short- and large-distance pictures.  Namely, if one
considers the lattice size $a\to 0$, then the corresponding coupling
$g(a)\to 0$ and the solution with a zero action \eq{leer} is strongly
favored at short distances.  At larger distances we are aware of the
dominance of the Abelian monopoles which have a non-zero nonabelian
action \cite{monact}. The end-points of a Dirac string still mark centers of
the Abelian monopole.  Thus, monopoles can be defined as point-like
objects topologically in terms of singular potentials, not action.

Similar logic holds in case of the so called P-vortices as well.
To detect the P-vortices one uses \cite{Deb97} the gauge 
maximizing the sum
\beq
\sum_{l}{|Tr~U_{l}|^2}
\eeq
where $l$ runs over all the links on the lattice.
The center projection is obtained by replacing
\beq
U_{l}~\rightarrow~sign~ (Tr~U_{l}).
\eeq
Each plaquette is marked either as $(+1)$ or $(-1)$
depending on the product of the signs assigned to the 
corresponding links.
P-vortex then pierces a plaquette with (-1). 
Moreover, the fraction $p$ of the total number of plaquettes pierced by the
P-vortices and of the total number of all the plaquettes $N_T$,
obeys the scaling law
\beq
p~=~{N_{vor}\over N_T}~\sim~f(\beta )
\eeq
where the function $f(\beta)$ is such that
$p$ scales like the string tension.
Assuming independence of the piercing for each plaquette
one has then for the center-projected Wilson loop $W_{cp}$:
\beq
W_{cp}=[(1-p)(+1)+p(-1)]^A~\approx~e^{-2pA}\label{tension}
\eeq
where $A$ is the number of plaquettes in the area stretched on
the Wilson loop. 
Numerically, Eq. (\ref{tension}) reproduces the full string tension.

It is quite obvious that the P-vortices defined this way correspond in
the continuum limit to singular gauge potentials $A_\mu^a$ (see, e.g.,
\cite{ShoStr2}. Indeed, the link matrices with the negative trace
correspond in the limit of the vanishing lattice spacing,
$a\to 0$ to the gauge potentials $A_l^a$ of the order $\frac 1
a$. Thus P-vortices correspond to large gauge potentials. The
potentials should mostly cancel, however, if the corresponding
field-strength tensors are calculated because of the asymptotic
freedom. The logic is essentially the same as outlined above for the
Dirac string, see, e.g.
\cite{tomboulis} and references therein.

At the moment, it is difficult to say a priori whether the topological
defects defined in terms of singular potentials can be
considered as infinitely thin from the physical point of view.
They might well be gauge artifacts.
Phenomenologically, using the topologically defined
point-like monopoles or infinitely thin
P-vortices one can measure non-perturbative $\bar{Q}Q$ potential 
at all the distances. 
It is remarkable therefore that the potentials
generated both by monopoles \cite{MonSh} and P-vortices \cite{Deb97}
turn to be linear at all the distances:
\beq
V_{non-pert}(r)~\approx~\sigma_{\infty} r ~\mbox{at all}~r\label{experiment}
\eeq 
Note that the Coulomb-like part is totally subtracted out through
the use of the topological defects. Moreover, no-change in the slope 
agrees well with the predictions of the dual Abelian Higgs model
(see Sect. 2).

The numerical observation 
\eq{experiment} is highly nontrivial.
If it were only the non-Abelian action that counts, 
then the non-perturbative
fluctuations labeled by the Dirac strings or by P-vortices
are bulky (see discussion above) and the corresponding 
$\bar{Q}Q$ potentials
\eq{experiment} should have been quadratic at short distances $r$.
This happens, for example, in the model \cite{greensite} with finite thickness 
of $Z_2$ vortices.
Also, if the lessons from the Georgi--Glashow model apply (see Sect 4
above) the finite size of the monopoles would spoil linearity of
the potential at short distances.

To summarize, direct measurements of the non-perturbative $\bar{Q}Q$
potential indicate the presence of a stringy potential at short distances.
The measurements go down to distances of order $(2GeV)^{-1}$.

\section{QCD phenomenology.}

In view of the results of measurements of the non-perturbative potential
it is interesting to reexamine \cite{ShoStr2}
the power corrections with the question in mind, whether there is
room for novel corrections associated with the short strings.
From the dimensional considerations alone it is clear
that the new corrections are of order $\sigma_0/Q^2$
where $Q$ is a large 
generic mass parameter characteristic for problem in hand.
Also, the ultraviolet renormalons in 4D indicate the same 
kind of correction.
Unlike the case of the non-perturbative potential
discussed above, other determinations of the power
corrections ask for a subtraction of the dominating perturbative
part. Which might make the results less definitive.
Here we briefly overview the relevant results.

({\it i}) The first claim of the existence of non-standard $1/Q^2$
corrections was made in ref. \cite{March}.
Namely, it was found that the
expectation value of the plaquette minus perturbation theory contribution
shows $1/Q^2$ behavior. On the other hand, the standard Operator Product
expansion results in a $1/Q^4$ correction.

{\it (ii)} The lattice simulation \cite{bali2} do not show any change
in the slope of the full $Q\bar{Q}$ potential as the distances are
changed from the largest to the smallest ones where Coulombic part
becomes dominant. It is known from phenomenological analysis and from
the calculations on the lattice \cite{Suz88,Blimit} that
the realistic QCD corresponds to the dual Abelian Higgs model with
$m_H \approx m_V$. As is mentioned in Sect.2, the AHM in the classical
approximation also gives $\sigma_\infty \approx \sigma_0$ at $m_H =
m_V$.

{\it (iii)} The explicit subtraction of the perturbative corrections
at small distances from $Q\bar{Q}$ potential in lattice gluodynamics
was performed in ref.\cite{bali3}. This procedure gives $\sigma_0
\approx 5 \sigma_\infty$ at very small distances.

{\it (iv)} There exist lattice measurements of fine splitting of the
$Q\bar{Q}$ levels as function of the heavy quark mass. The
Voloshin-Leutwyler \cite{VL} picture predicts a particular pattern of
the heavy mass dependence of this splitting. Moreover, these predictions are
very different from the predictions based on adding a linear part to
the Coulomb potential (Buchmuller-Tye potential \cite{BuTy81}).  The
results of recent calculations of this type \cite{Fi98} favor
the linear correction to the
potential at short distances.

({\it v}) Analytical studies of the Bethe-Salpeter equation and comparison
of the results with the charmonium spectrum data favor a
non-vanishing linear correction to the potential at short distances
\cite{BaMo99,Faustov}.

({\it vi}) The lattice-measured instanton density 
as a function of the instanton
size $\rho$ does not satisfy
the standard OPE predictions that the leading correction is
of order $\rho^4$.
Instead, the leading corrections is in fact quadratic  \cite{shuryak}.

({\it vii}) One of the most interesting manifestation of short strings might be
$1/Q^2$ corrections to the standard OPE for current--current
correlation function $\Pi(Q^2)$. It is
impossible to calculate the coefficient of $1/Q^2$ corrections
from first principles, and in
ref. \cite{CNZ99} it was suggested to simulate this correction by a
tachyonic gluon mass. The Yukawa potential with an imaginary mass has
the linear attractive piece at small distances, i.e. reproduces
short strings. The use of the gluon propagator with the imaginary
gluon mass ($m^2_g = -0.5 \mbox{ Gev}^2$) explains unexpectedly well 
the behavior of $\Pi(Q^2)$ in various channels. To check the model
with a tachyonic short-distance mass further, it would be very
important to perform the accurate calculations of various correlators
$\Pi(Q^2)$ on the lattice. There are also alternative
theoretical schemes in QCD \cite{ImT} which predict 
non-conventional $1/Q^2$.

\section{Concluding Remarks}

As is revealed by analysis of the data on the power corrections,
the existence of the novel quadratic corrections 
is strongly supported by the data. 
There are, however, 
two caveats to the statement that the novel short-distance 
power corrections have been detected. On the theoretical side, 
the existence of short strings has been proven in 
only within the Abelian Higgs model. As for the QCD itself,
the analysis is so far inconclusive.
On the experimental side, the data always refer to a limited range of
distances. In particular linear non-perturbative potential has been
observed at distances of order of one lattice spacing which
is in physical units is $(1 \div 2 ~GeV)^{-1}$. 
One can argue that at shorter distances the
behavior of the non-perturbative power corrections
changes (see, e.g., \cite{huber,shuryak}).
Which would be a remarkable phenomenon by itself. 

\section*{Acknowledgments}

M.N.Ch. and M.I.P. acknowledge the kind hospitality of the staff of
the Max-Planck Institut f\"ur Physik (M\"unchen), where the part of
this work has been done.  Work of M.N.C., F.V.G. and M.I.P.  was
partially supported by grants RFBR 99-01230a and INTAS 96-370.

\end{document}